\documentclass[a4paper,11pt]{article}
\usepackage{jheppub} 

\usepackage{easyReview}
\newcommand{\e}{\epsilon}
\renewcommand{\a}{\alpha}

\graphicspath{{Figs/}}
\title{\boldmath Method of regions for dual conformal integrals}

\author[a]{Leonid V. Bork,}
\author[b,c,1]{Roman N. Lee\note{Corresponding author.},}
\author[c]{Andrei I. Onishchenko}
\affiliation[a]{The Center for Fundamental and Applied Research, 127030 Moscow, Russia}
\affiliation[b]{Budker Institute of Nuclear Physics,
630090 Novosibirsk, Russia}
\affiliation[c]{Joint Institute for Nuclear Research, 141980 Dubna, Russia}

\emailAdd{r.n.lee@inp.nsk.su}

\abstract{
We apply the method of regions to the evaluation of dual conformal integrals with small off-shellness. In contrast to conventional approach, where the separation of regions is performed via dimensional regularization breaking the dual conformal invariance (DCI), we use a sufficiently generic combination of dimensional and analytic regularizations which preserves the DCI. Within this regularization (dubbed as DCI regularization), the contribution of each region becomes DCI. We show that our method dramatically simplifies the calculations.  As a demonstration, we calculate the slightly off-shell DCI pentabox integral up to power corrections. The contributions of all 32 regions appear to be expressible in terms of products/ratios of $\Gamma$-functions multiplied by some powers of DCI cross-ratios. Therefore, after removing the regularization, we obtain the final expression in terms of cross-ratios logarithms only. We have checked that our result for pentabox integral numerically agrees with the result of the recent Belitsky\&Smirnov paper \cite{Belitsky:2025sin} which has essentially more complicated form.}

\begin{document}
\maketitle
\flushbottom
\section{Introduction}
Dual conformal symmetry \cite{ConformalProperties4point,Drummond:2008vq} is essential for our ability to obtain exact analytical results in N=4 SYM, which, in turn, is best available playground for the investigation of perturbative QCD in particular and general properties of the four dimensional quantum field theories in general. Usually for the scattering amplitudes in N=4 SYM the dual conformal symmetry is broken due to the presence of the IR divergences although in somewhat controlled fashion \cite{Drummond:2007au}. Despite that, one can find kinematic regimes \cite{HennGiggs1,Caron-Huot:2021usw} or formulation of the problem \cite{Coronado:2018cxj,Bercini:2020msp,Caron-Huot:2021usw} where dual conformal symmetry remains intact. To test different integrability-based all-loop conjectures or bootstrap results for such regimes, explicit perturbative computations are required. This boils down to our ability to evaluate dual conformal multiloop integrals in various asymptotic kinematic regimes \cite{Coronado:2018cxj,Dectagon,Dectagon2,Bork:2022vat}. One of such regimes, which turned out to be interesting \cite{Coronado:2018cxj,Bercini:2020msp,Caron-Huot:2021usw,Dectagon2,Bork:2022vat}, corresponds to the situation when all or some of the dual conformal cross ratios are small. In terms of usual kinematic variables, these asymptotic regimes correspond to the condition $p_i^2=m^2\ll s_j$. Hence we will call such an asymptotic regime \emph{slightly off-shell}.

The main tool for the calculation of the asymptotics of multiloop integrals is the method of regions (MofR) \cite{Beneke:1997zp,Pak:2010pt,Semenova:2018cwy} which allows one to separate the contributions of different integration regions. In this method, the contribution of each region is represented by an integral obtained from the original integral by rescaling the integration variables and expanding under the integral sign. For this method to work, one usually needs dimensional regularization, even when the original integral is convergent. In more complicated cases, when the dimensional regularization fails to separate the regions, one has to resort to auxiliary analytic regularization. 
For the calculations of Lorentz-covariant multiloop integrals, the dimensional regularization proves to be the most convenient as it retains the Lorentz invariance. In particular, the scalar integrals in dimensional (and analytic) regularization still depend on the external momenta only via the kinematic invariants --- invariants with respect to Lorentz transformations.

In the context of dual conformal integrals, there is one drawback of these regularizations. Although they preserve the Poincare invariance in dual coordinates, they generally break the invariance with respect to the inversion. Thus, the results obtained within dimensional regularization recover their dual conformal invariance only after the regularization is removed.

However, it is, in principle, possible to introduce a combination of dimensional and analytic regularization such that the regularized integrals remain dual conformal invariant. In fact, as we shall see, this combination is not unique, but depends on $I-L$ additional parameters, where $I$ is the number of internal lines and $L$ is the number of loops. For divergent integrals this ambiguity represents a conceptual problem. However, for the calculation of asymptotics of convergent integrals via MofR it provides an additional lever to separate the contribution of regions.

In the present paper, we demonstrate that the described approach drastically simplifies the calculation of the slightly off-shell asymptotics of DCI integrals at small off-shellness. In several considered examples, we have successfully expressed the contribution of each region in terms of products of $\Gamma$-functions exactly in all parameters of DCI-conserving regularization.


\section{DCI integrals in \texorpdfstring{$d$}{d} dimensions}
Let us consider the $P$-point $L$-loop DCI integral in $4$ dimensions\footnote{All through the paper we use Euclidean metrics.}
\begin{equation}
    I_{L}(y_1,\ldots y_P)=
    \int \prod_{l=P+1}^{P+L}\frac{d^4y_{l}}{\pi^2}
    \prod_{i=1}^{M} \left[(y_{k_i}-y_{m_i})^2\right]^{-n_i}.
\end{equation}
where $n_i$ are integers and $y_i$ are standard dual variables. 
The Poincare invariance of this integral is obvious, but the dual conformal invariance also requires that the integral is invariant with respect to the inversion $y_n\to y_n/y_n^2$. In order to derive the precise condition on the indices, let us introduce the indicator function
\begin{equation}
\theta_{li}=
    \begin{cases}
        1 & \text{if }l\in\{k_i,m_i\}\\
        0 & \text{otherwise}
    \end{cases}
\end{equation}
Then, under inversion the integrand acquires the factor
\begin{equation}
    \prod_{l=1}^{P} \left(y_l^2\right)^{\sum_in_i\theta_{li}}
    \prod_{l=P+1}^{P+L} \left(y_l^2\right)^{-4+\sum_in_i\theta_{li}}
\end{equation}
Thus, the integral is DCI invariant iff
\begin{equation}\label{eq:dci_constraint_4}
    \sum_in_i\theta_{li}=
    \begin{cases}
        0,&l\leqslant P\\
        4,& l>P
    \end{cases}
\end{equation}

Let us now regularize the integral, both dimensionally and analytically, such that it remains DCI invariant. 
In addition to the dimensional regularization parameter $\e=2-d/2$, we introduce the parameters $\a_1,\ldots, \a_M$, so that the regularized integral reads 
\begin{equation}\label{eq:Ireg}
    I_{L}^{\text{reg}}(y_1,\ldots y_P)=
    \int \prod_{l=1}^{L}\frac{d^dy_{P+l}}{\pi^{d/2}}
    \prod_{i=1}^{M} \left[(y_{k_i}-y_{m_i})^2\right]^{-\nu_i},
\end{equation}
where $\nu_i=n_i+\a_i$. Let us now discuss the restrictions on the choice of $\a_i$. A technical requirement is that we don't want to introduce analytic regularization for numerators which depend on integration variables. In other words, we put $\a_i=0$ if $\max(k_i,m_i)>P$ and $n_i\leqslant 0$. Next, we have to request the invariance with respect to inversion. This restriction reads
\begin{equation}\label{eq:dci_constraint_d}
    \sum_i\nu_i\theta_{li}=
    \begin{cases}
        0,&l\leqslant P\\
        d,& l>P
    \end{cases}
\end{equation}
Subtracting \eqref{eq:dci_constraint_4} from \eqref{eq:dci_constraint_d}, we obtain the conditions for the regularization parameters 
\begin{equation}\label{eq:dci_constraint_d1}
    \sum_i\a_i\theta_{li}=
    \begin{cases}
        0,&l\leqslant P\\
        -2\e,& l>P
    \end{cases}
\end{equation}
In particular, the above equations for $l>P$ make obvious the fact that one cannot avoid analytic regularization for internal lines, since in this case the sum in the left-hand side is zero. 
Although analytical regularization is expected to complicate traditional approaches, including IBP reduction, we will see below that the benefits of retaining DCI symmetry outweigh these complications.

\section{Pedagogical example: slightly off-shell box}
Let us consider the one-loop massless box integral \cite{Usyukina:1992jd,Usyukina:1993ch} in $4$ dimensions with slightly off-shell legs. In dual coordinates it reads 
\begin{equation}\label{eq:box4}
    B =\int \frac{d^4y_5}{\pi^{2}} \frac{y_{13}^2y_{24}^2}{y_{15}^2y_{25}^2y_{35}^2y_{45}^2},
\end{equation}
where we use the notation $y_{ij}=y_i-y_j$.
We define kinematic invariants 
\begin{gather}
    p_1^2=y_{14}^2,\quad
    p_2^2=y_{21}^2,\quad
    p_3^2=y_{32}^2,\quad
    p_4^2=y_{43}^2,\nonumber\\
    s=(p_2+p_1)^2=y_{24}^2,\quad
    t=(p_3+p_2)^2=y_{13}^2.
\end{gather}
Note the factor $y_{13}^2y_{24}^2=st$ in the definition of $B$, Eq. \eqref{eq:box4}, which makes $B$ dual conformal invariant, so that it depends on the kinematic invariants only via 2 DCI cross-ratios
\begin{equation}
    u_1=\frac{p_1^2p_3^2}{st},\quad u_2=\frac{p_2^2p_4^2}{st}.
\end{equation}
Of course, thanks to Davydychev\&Ussyukina 
\cite{Usyukina:1992jd,Usyukina:1993ch}, we can evaluate this integral for generic invariants. 

But let us pretend that we don't know this result and are interested in the limit of small off-shellness, $p_i^2\ll s,t$. Then we can rely on the method of regions \cite{Beneke:1997zp,Pak:2010pt,Semenova:2018cwy}. This method requires some sort of regularization, which allows one to separate the contribution of each region. The conventional choice is the dimensional regularization. 

Let us consider the most generic regularization, which also includes the analytic regularization of all propagators:
\begin{equation}\label{eq:box_d}
    B^{\text{gen}} =\int \frac{d^dy_5}{\pi^{d/2}} 
    \frac{\left[y_{13}^2\right]^{1-\a_5}\left[y_{24}^2\right]^{1-\a_6}
    \left[y_{21}^2\right]^{-\a_7}\left[y_{32}^2\right]^{-\a_8}}{\left[y_{15}^2\right]^{1+\a_1}\left[y_{25}^2\right]^{1+\a_2}\left[y_{35}^2\right]^{1+\a_3}\left[y_{45}^2\right]^{1+\a_4}},
\end{equation}

The Feynman parametric representation reads
\begin{multline}
    B^{\text{gen}}=\frac{t^{1-\a_5} s^{1-\a_6}}{\left(p_2^2\right)^{\a_7} \left(p_3^2\right)^{\a_8}} \Gamma \left(2+\a_{1234}+\e \right)
    \int \frac{ dx_1x_1^{\a_1} dx_2x_2^{\a_2} dx_3x_3^{\a_3} dx_4x_4^{\a_4}}{\Gamma \left(1+\a_1,1+\a_2,1+\a_3,1+\a_4\right)}\\
    \times
    \frac{\delta\left(1-\sum x\right)x_{1234}^{\a_{1234}+2 \e } }{\left(
    \hat{x}_{24} s+\hat{x}_{13} t+\hat{x}_{14} p_1^2+\hat{x}_{12} p_2^2+\hat{x}_{23} p_3^2+\hat{x}_{34}p_4^2
    \right)^{2+\a_{1234}+\e}}\,,
\end{multline}
where we used the notations $x_{ij\ldots k}=x_i+x_j+\ldots+ x_k$, $\a_{ij\ldots k}=\a_i+\a_j+\ldots+ \a_k$, and $\hat{x}_{ij}=x_ix_j$.

The conventional dimensional regularization corresponds to setting all $\a_i$ to zero. Let us now apply the constraints \eqref{eq:dci_constraint_d1} which ensure that the integral is DCI. They read
\begin{gather}\label{eq:alpha_box}
    \a_4+\a_6=0,\quad
    \a_1+\a_5+\a_7=0,\quad
    \a_3+\a_5+\a_8=0,\quad
    \a_2+\a_6+\a_7+\a_8=0,\nonumber\\
    \a_1+\a_2+\a_3+\a_4=-2 \e.\label{eq:dci_const}
\end{gather}
These equations can be used to express, for example, $\a_{4-8}$ via $\a_{1-3}$.

\begin{figure}
    \centering
    \includegraphics[width=0.5\linewidth]{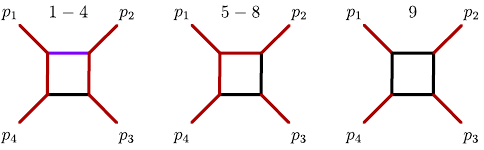}
    \caption{Regions, contributing to box diagram at small off-shellness. Black, red and magenta lines correspond to hard, soft and ultrasoft propagators. Each of the two first diagrams depicts four regions obtained by its rotations.}
    \label{fig:box_regions}
\end{figure}

Using the approach of Refs. \cite{Pak:2010pt,Semenova:2018cwy},\footnote{We use a custom implementation \texttt{GetRegions} (\url{https://github.com/rnlg/Get-regions}) of the approach of Refs. \cite{Pak:2010pt,Semenova:2018cwy}.} we discover 9 regions, graphically depicted in Fig. \ref{fig:box_regions}. For example, the first region corresponds to $[x_1:x_2:x_3:x_4] \sim [m^{-4}:m^{-2}:1:m^{-2}]$ scaling, where $m^2\sim p_i^2$ denotes the characteristic scale of off-shellness.

We denote the contribution of $k$-th region as $B_k^{\text{dim}}$ and $B_k^{\text{dci}}$ for dimensional regularization and DCI regularization, respectively. These contributions in the leading order are shown in Table \ref{tab:box_regions}.

\renewcommand{\arraystretch}{1.75}
\begin{table}[h]
    \centering
    \begin{tabular}{|c|c|l|}
    \hline
     $k$ & $B_k^{\text{dim}}$ & $B_k^{\text{dci}}$
    \\\hline
      1 &  
        $s^{\e } \Gamma (1-\e ) \Gamma (\e )^2 \left(p_1^2 p_2^2\right)^{-\e }$ &
        $\frac{\Gamma \left(1-\a_1-\e, \a_1+\a_2+\e,\a_1+\a_4+\e \right) }{\Gamma \left(1+\a_1,1+\a_2,1+\a_4\right)}u_1^{\a_2+\a_3+\e }$
    \\\hline
      2 &  
        $t^{\e } \Gamma (1-\e ) \Gamma (\e )^2 \left(p_2^2 p_3^2\right)^{-\e }$ &
        $\frac{\Gamma \left(1-\a_2-\e,\a_1+\a_2+\e,\a_2+\a_3+\e \right)}{\Gamma \left(1+\a_1,1+\a_2,1+\a_3\right)}$
    \\\hline
      3 &  
        $s^{\e } \Gamma (1-\e ) \Gamma (\e )^2 \left(p_3^2 p_4^2\right)^{-\e }$ &
        $\frac{\Gamma \left(1-\a_3-\e,\a_3+\a_4+\e,\a_2+\a_3+\e \right) }{\Gamma \left(1+\a_2,1+\a_3,1+\a_4\right)}u_2^{\a_1+\a_2+\e }$
    \\\hline
      4 &  
        $t^{\e } \Gamma (1-\e ) \Gamma (\e )^2 \left( p_4^2p_1^2\right)^{-\e }$ &
        $\frac{\Gamma \left(1-\a_4-\e,\a_3+\a_4+\e, \a_1+\a_4+\e \right) }{\Gamma \left(1+\a_1,1+\a_3,1+\a_4\right)} u_1^{\a_2+\a_3+\e }u_2^{\a_1+\a_2+\e }$
    \\\hline
      5 &  
        ${\Gamma (-\e )^2 \Gamma (\e ) \left(p_1^2\right)^{-\e }}/{\Gamma (-2 \e )}$ &
        $0$
    \\\hline
      6 &  
        ${\Gamma (-\e )^2 \Gamma (\e ) \left(p_2^2\right)^{-\e }}/{\Gamma (-2 \e )}$ &
        $0$
    \\\hline
      7 &  
        ${\Gamma (-\e )^2 \Gamma (\e ) \left(p_3^2\right)^{-\e }}/{\Gamma (-2 \e )}$ &
        $0$
    \\\hline
      8 &  
        ${\Gamma (-\e )^2 \Gamma (\e ) \left(p_4^2\right)^{-\e }}/{\Gamma (-2 \e )}$ &
        $0$
    \\\hline
      9 &  
        Eq. \eqref{eq:B9dim} &
        $0$
    \\\hline
    \end{tabular}
    \caption{Contribution of different regions in dimensional and DCI regularizations. Note that in the second column $\a_1, \a_2,\a_3,\a_4$ are assumed to satisfy Eq. \eqref{eq:dci_const}.}
    \label{tab:box_regions}
\end{table}
In particular, the ``hard'' contribution $B_9$ of the 9th region in dimensional regularization reads
\begin{multline}\label{eq:B9dim}
    B_9^{\text{dim}}=\frac{st\pi  \Gamma(-\e)}{\sin (\pi  \e )\Gamma (-2 \e ) }
     \bigg[
        2 \pi \cot (\pi  \e ) \left(\tfrac{s+t}{st}\right)^{\e }\\
        -\tfrac{t s^{-1-\e}}{1+\e} {}_2F_1\left(1,1;2+\e;-\tfrac{t}{s}\right)
        -\tfrac{s t^{-1-\e}}{1+\e} {}_2F_1\left(1,1;2+\e;-\tfrac{s}{t}\right)\bigg]
\end{multline}
There are several striking differences between the results of the two regularizations. First, each contribution in DCI regularization is a function of dual conformal invariant cross-ratios only, as expected. Second, many contributions in the DCI regularization vanish. It is especially remarkable for the region 9, where the result of dimensional regularization can not be expressed in terms of $\Gamma$-functions.
Although the contributions of individual regions are different, their sums in the limit $\e,\a_i\to0$ should coincide for both regularizations. Indeed, in both cases we find
\begin{equation}
    B = \log u_1 \log u_2 +2\zeta_2 + O (u_1,u_2)\,.
\end{equation}

\section{Slightly off-shell pentabox}
\begin{figure}
    \centering
    \includegraphics[width=0.7\linewidth]{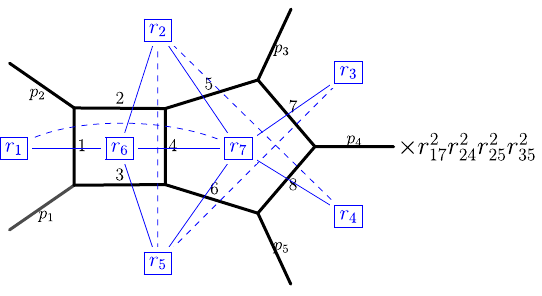}
    \caption{DCI pentabox integral. On the dual graph (shown in blue) the dashed lines denote numerators.}
    \label{fig:pentabox}
\end{figure}

It is remarkable that exactly the same approach works perfectly for the calculation of slightly off-shell two-loop dual conformal pentabox integral given by the graph shown in Fig. \ref{fig:pentabox}. 

The most generic DCI regularization reads
\begin{multline}
    PB=PB(\e,\a_1,\ldots, \a_6)\\
    =\int \frac{d^dr_6}{\pi^{d/2}}\frac{d^dr_7}{\pi^{d/2}}
    \frac{r_{14}^{2 \left(\a_3-\a_{246}\right)} r_{25}^{2 \left(1+\a_{45}\right)} r_{31}^{2 \left(\a_{1246}-\a_3\right)} r_{42}^{2 \left(1+\a_2-\a_5\right)} r_{53}^{2 \left(1+\a_3-\a_{124}\right)}r_{17}^2}{r_{56}^{2 \left(1+\a_3-\a_{12}\right)} r_{57}^{2 \left(1+\a_5\right)} r_{61}^{2 \left(1+\a_1\right)} r_{62}^{2 \left(1+\a_2\right)} r_{72}^{2 \left(1+\a_4\right)} r_{73}^{2 \left(1+\a_6\right)} r_{74}^{2 \left(1+\a_3-\a_{456}\right)} r_{67}^{2 \left(1-\a_3-2\e \right)}}
\end{multline}
 Due to dual conformal symmetry the integral depends on external momenta only via five cross-ratios, which we choose as\footnote{As usual, all indices are understood $\mod (5)$, e.g. $6=1 \mod(5)$.}
 \begin{equation}
     u_i=\frac{r_{i+1,i+2}^2r_{i-2,i-1}^2}{r_{i+2,i-2}^2r_{i+1,i-1}^2}=\frac{p_{i+2}^2p_{i-1}^2}{p_{i-2}^2s_i},
 \end{equation}
 where $p_i^2=(r_i-r_{i-1})^2, s_i=(r_{i+1}-r_{i-1})^2=(p_{i+1}+p_i)^2$ are conventional kinematic variables. We are interested in the asymptotics 
$p_i^2=m^2\ll s_{j}$, when 
\begin{equation}
    u_i=m^2/s_i\ll 1.
\end{equation}
 
We use Lee-Pomeransky parametric representation \cite{Lee:2013hzt,Lee:2014tja} which gives
\begin{multline}
    PB=s_1^{1+\a_{45}} s_2^{\a_{12\bar 346}} s_3^{1+\a_2-\a_5} s_4^{1-\a_{12\bar 34}} s_5^{-\a_{2\bar 3 4 6}} \intop_{\mathbb{R}_+^8} dx_1\ldots dx_8\\
    \times\frac{ \Gamma (3-\epsilon ) x_1^{\a_1} x_2^{\a_2} x_3^{-\a_{12\bar 3}} x_4^{-\a_3-2\e }x_5^{\a_4} x_6^{\a_5} x_7^{\a_6} x_8^{-\a_{\bar 3456}} G^{-3+\epsilon}}{\Gamma \left(1+\a_1,1+\a_2,1-\a_{12\bar 3},1+\a_4,1+\a_5,1-\a_{\bar 3456},1+\a_6,1+\a_{\bar 3}-2\e,\a_{\bar 3}-1-\epsilon\right)}\\
    \times\bigg(m^2 \left(x_1x_{2356}+x_{23}x_{456}+x_4x_{56}\right)    +s_1 \hat{x}_{23}+s_2 x_{1234} x_7+s_5 x_{1234} x_8+x_{1234}\bigg)
    \label{eq:pentabox_lp}
\end{multline}
where
\begin{multline}
    G=x_{1234} x_{5678}+x_{123} x_4+s_1 \left(x_{123}\hat{x}_{56}+\hat{x}_{23}x_{5678}+x_{25} x_{36} x_4\right)\\
    +s_2 \hat{x}_{147}
    +s_3 \left(x_{1234} \hat{x}_{58}+ \hat{x}_{248}\right)
    +s_4 \left(x_{1234} \hat{x}_{67}+ \hat{x}_{347}\right) +s_5 \hat{x}_{148}\\
    +m^2 \big(\hat{x}_{14}x_{2356}+\hat{x}_{247}+\hat{x}_{348}+x_1x_{23}x_{5678}+x_{1234}\hat{x}_{57}+x_{1234}x_{67}x_8\big)   
\end{multline}
Here we use notations $x_{ij\ldots k}=x_i+x_j+\ldots+ x_k$, $\hat{x}_{ij\ldots k}=x_i\cdot x_j\cdot\ldots\cdot x_k$, $\a_{ij\ldots k}=\a_i+\a_j+\ldots+ \a_k$, and $\a_{\bar n}=-\a_n$.

\begin{figure}
    \centering
    \includegraphics[width=1\linewidth]{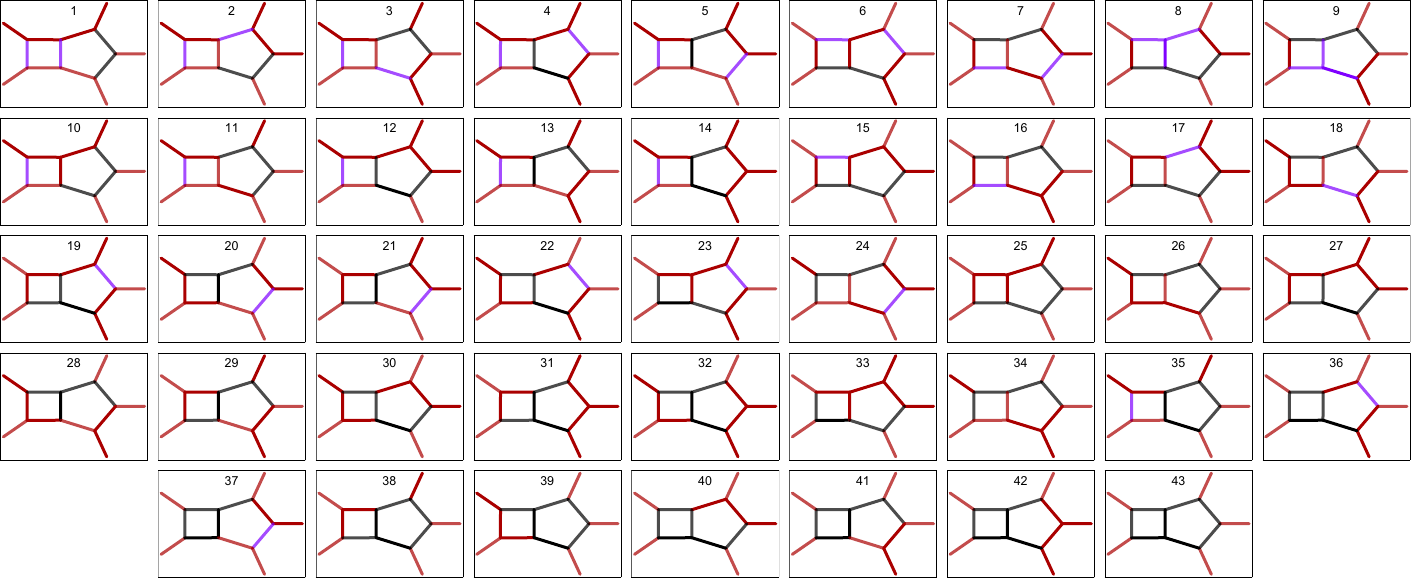}
    \caption{43 regions which contribute to small-$m^2$ asymptotics.}
    \label{fig:pentabox_regions}
\end{figure}

Using \texttt{GetRegions} package, we find 43 regions depicted in Fig. \ref{fig:pentabox_regions}. Regions \#\#1, 2, 3, 8, 9, 10, 11, 17, 18, 25, 26 do not contribute in the leading order. The contributions of other regions, except the hard region \#43 can be obtained by explicitly integrating their parametric representation using the only master formula
\begin{equation}
    \intop_0^{\infty} dx \frac{x^{\alpha-1}}{(A+B x)^{\beta}}=\frac{\Gamma(\a)\Gamma(\beta-\a)}{\Gamma(\beta)}A^{\alpha-\beta} B^{-\alpha}\,.
\end{equation}
As a result, the contributions of all regions is expressed as a product of $\Gamma$-functions multiplied by some powers of invariant cross-ratios.

As an example, consider region \#38, which seems to be one of the most complicated regions (apart from the hard region \#43) when considered in conventional dimensional regularization.
After rescaling $x_{1,2}\to m^{-2}x_{1,2}$ in Eq. \eqref{eq:pentabox_lp} and retaining only the leading term in $m$ in the integrand, we obtain
\begin{multline}
    PB_{38}^{\text{dci}}=m^{2 \left(\a_{\bar 345}-\epsilon \right)}\Gamma (3-\epsilon )s_1^{1+\a_{45}} s_2^{\a_{12\bar 346}} s_3^{1+\a_{2\bar 5}} s_4^{1-\a_{12\bar 34}} s_5^{-\a_{2\bar 346}} \intop_{\mathbb{R}_+^8} dx_1\ldots dx_8\\
    \times \frac{x_1^{\a_1} x_2^{\a_2} x_3^{-\a_{12\bar 3}} x_4^{-\a_3-2\e } x_5^{\a_4} x_6^{\a_5} x_7^{\a_6} x_8^{-\a_{\bar 3456}}}{\Gamma \left(1+\a_1,1+\a_2,1-\a_{12\bar 3},1+\a_4,1+\a_5,1-\a_{\bar 3456},1+\a_6,1+\a_{\bar 3}-2\e ,\a_{\bar 3} -1 -\epsilon\right)} \\
    \times\frac{  x_{1234}(s_2  x_7+s_5 x_8)}{\left(x_{1234} (x_{78}+\hat{x}_{78}+s_3 \hat{x}_{58}+s_4 \hat{x}_{67})+s_1 \hat{x}_{23} x_{78}
    +s_2 \hat{x}_{147}
    +s_3 \hat{x}_{248}
    +s_4 \hat{x}_{347}
    +s_5 \hat{x}_{148}\right)^{3-\epsilon}}\,,
\end{multline}
where we use notation $\Gamma(a_1,\ldots, a_k)=\Gamma(a_1)\cdot\ldots \cdot \Gamma(a_k)$.
We make subsequent rescalings 
\begin{align}
     x_{5,6,7,8} \to x_4 x_{5,6,7,8}, 
     \quad 
     x_8 \to x_6 x_8,
     \quad
     x_2\to x_1 x_2\,.
\end{align}
and take all the integrals in the following sequence: $x_3,x_4,x_5,x_7,x_6,x_1,x_2,x_8$.
Then we obtain
\begin{multline}
    PB_{38}^{\text{dci}}=-\left(\frac{u_4}{u_1u_2}\right)^{\alpha_{12}+\epsilon }\\
    \times
    \frac{\left(\alpha _2+\epsilon \right) \Gamma \left(1+\epsilon +\alpha _{3\bar4},1+\alpha _{2\bar5},-\alpha _{2\bar346},\alpha _{2\bar3}-\epsilon ,\alpha _{\bar345}-\epsilon ,\alpha _{\bar346}-\epsilon ,\alpha _{12}+\epsilon \right)}{\Gamma \left(1-\epsilon +\alpha _{2\bar34},1+\alpha _1,1+\alpha _2,1+\alpha _5,1-\alpha _{\bar3456},1+\alpha _6,1-2 \epsilon -\alpha _3\right)}\,.
\end{multline}

\renewcommand{\arraystretch}{1.2}
\begin{table}[h!]
    \centering
    \begin{tabular}{|c|l|l|}
    \hline
    \# & rescalings & sequence\\\hline
    $4$ & $$ & $x_1, x_2, x_3, x_4, x_5, x_6, x_7, x_8$\\\hline
    $5$ & $$ & $x_1, x_2, x_4, x_3, x_5, x_8, x_6, x_7$\\\hline
    $6$ & $$ & $x_1, x_3, x_2, x_4, x_6, x_5, x_7, x_8$\\\hline
    $7$ & $$ & $x_1, x_2, x_3, x_4, x_5, x_6, x_8, x_7$\\\hline
    $12$ & $x_7\to x_5 x_7$ & $x_1, x_2, x_4, x_3, x_6, x_8, x_5, x_7$\\\hline
    $13$ & $x_8\to x_6 x_8$ & $x_1, x_2, x_3, x_5, x_7, x_4, x_6, x_8$\\\hline
    $14$ & $x_8\to x_7 x_8$ & $x_1, x_2, x_3, x_5, x_6, x_4, x_7, x_8$\\\hline
    $15$ & $x_{5,7}\to x_4 x_{5,7}$ & $x_1, x_3, x_6, x_8, x_2, x_4, x_7, x_5$\\\hline
    $16$ & $x_{6,8}\to x_4 x_{6,8}$ & $x_1, x_2, x_5, x_3, x_7, x_4, x_8, x_6$\\\hline
    $19$ & $x_2\to x_1 x_2$ & $x_3, x_4, x_5, x_6, x_7, x_8, x_1, x_2$\\\hline
    $20$ & $x_3\to x_1 x_3$ & $x_2, x_4, x_5, x_6, x_8, x_7, x_1, x_3$\\\hline
    $21$ & $x_2\to x_1 x_2$ & $x_3, x_4, x_6, x_5, x_8, x_7, x_1, x_2$\\\hline
    $22$ & $x_3\to x_1 x_3$ & $x_2, x_4, x_5, x_6, x_7, x_8, x_1, x_3$\\\hline
    $23$ & $x_4\to x_2 x_4,\ x_8\to x_7 x_8$ & $x_1, x_3, x_5, x_6, x_2, x_7, x_4, x_8$\\\hline
    $24$ & $x_4\to x_3 x_4,\ x_8\to x_7 x_8$ & $x_1, x_2, x_5, x_6, x_7, x_3, x_8, x_4$\\\hline
    $27$ & $x_2\to x_1 x_2,\ x_7\to x_5 x_7$ & $x_3, x_4, x_6, x_8, x_1, x_5, x_7, x_2$\\\hline
    $28$ & $x_3\to x_1 x_3,\ x_8\to x_6 x_8$ & $x_2, x_4, x_5, x_7, x_1, x_6, x_8, x_3$\\\hline
    $29$ & $x_2\to x_1 x_2,\ x_8\to x_6 x_8$ & $x_3, x_4, x_5, x_7, x_1, x_6, x_2, x_8$\\\hline
    $30$ & $x_3\to x_1 x_3,\ x_7\to x_5 x_7$ & $x_2, x_6, x_8, x_4, x_1, x_5, x_3, x_7$\\\hline
    $31$ & $x_2\to x_1 x_2,\ x_8\to x_7 x_8$ & $x_3, x_4, x_5, x_6, x_1, x_7, x_2, x_8$\\\hline
    $32$ & $x_3\to x_1 x_3,\ x_8\to x_7 x_8$ & $x_2, x_4, x_5, x_6, x_1, x_7, x_3, x_8$\\\hline
    $33$ & $x_{4,5,7}\to x_2 x_{4,5,7},\ x_{5,7}\to \frac{x_4 x_{5,7}}{x_4+1}$ & $x_1, x_8, x_3, x_6, x_2, x_4, x_7, x_5$\\\hline
    $34$ & $x_{4,6,8}\to x_3 x_{4,6,8},\ x_{6,8}\to \frac{x_4 x_{6,8}}{x_4+1}$ & $x_1, x_2, x_5, x_7, x_3, x_4, x_8, x_6$\\\hline
    $35$ & $x_{5,6,7,8}\to x_4 x_{5,6,7,8}$ & $x_1, x_2, x_3, x_4, x_5, x_7, x_8, x_6$\\\hline
    $36$ & $x_{2,3,4}\to x_1 x_{2,3,4}$ & $x_5, x_6, x_7, x_8, x_1, x_4, x_3, x_2$\\\hline
    $37$ & $x_{2,3,4}\to x_1 x_{2,3,4}$ & $x_5, x_6, x_8, x_7, x_1, x_4, x_2, x_3$\\\hline
    $38$ & $x_{5,6,7,8}\to x_4 x_{5,6,7,8},\ x_8\to x_6 x_8,\ x_2\to x_1 x_2$ & $x_3, x_4, x_5, x_7, x_6, x_1, x_2, x_8$\\\hline
    $39$ & $x_{5,6,7,8}\to x_4 x_{5,6,7,8},\ x_7\to x_5 x_7,\ x_3\to x_1 x_3$ & $x_2, x_4, x_6, x_8, x_5, x_1, x_3, x_7$\\\hline
    $40$ & $x_{2,3,4}\to x_1 x_{2,3,4},\ x_{5,7}\to x_4 x_{5,7}$ & $x_8, x_6, x_1, x_2, x_3, x_4, x_7, x_5$\\\hline
    $41$ & $x_{2,3,4}\to x_1 x_{2,3,4},\ x_{6,8}\to x_4 x_{6,8}$ & $x_7, x_5, x_1, x_3, x_2, x_4, x_8, x_6$\\\hline
    $42$ & $x_{2,3,4}\to x_1 x_{2,3,4},\ x_8\to x_7 x_8$ & $x_5, x_6, x_1, x_2, x_4, x_3, x_7, x_8$\\\hline
    \end{tabular}
    \caption{Rescaling of the integration variables and order of integration. The last integration in the regions \#\#19, 20, 27, 28, 36, 37, 40, 41 gives a scaleless integral equal to zero.}
    \label{tab:scale_seq}
\end{table}
In table \ref{tab:scale_seq} we present the rescaling and order of integration for each region that contribute in the leading order in $m$, except the last region \#43. The last integration in regions \#\#19, 20, 27, 28, 36, 37, 40, 41 is scaleless, therefore, their contributions are zero.

As to this last region \#43, we proceed as follows. The contribution of this region is obtained by putting $m=0$ in the integrand of Eq. \eqref{eq:pentabox_lp}. However, it is still difficult to find the appropriate change of variables that allows for direct integration. Instead, we adopt another strategy. We use the fact that this contribution is dual conformal invariant (as well as others) and does not depend on $m^2$. As it is not possible to construct a nontrivial DCI invariant with $s_i$ only, this contribution should be equal to a constant. Thus, we can consider consecutive limits $s_i\to 0$, and for each limit keep only the leading contribution proportional to $s_i^0$. We choose the following sequence of limits
\begin{equation}
    PB_{43}^{\text{dci}}=\lim_{s_4\to 0}\lim_{s_2\to 0}\lim_{s_5\to 0}\lim_{s_3\to 0}\lim_{s_1\to 0}PB_{43}^{\text{dci}}.
\end{equation}
In the asymptotics $s_1\to 0$ we find 3 distinct regions, but only one of them gives the term $\propto s_1^0$. This region corresponds to the rescaling 
\begin{equation}
    x_{1,2,3,4,5,6}\to s_1^{-1} x_{1,2,3,4,5,6}\,.
\end{equation}
Performing this rescaling and taking the limit $s_1\to 0$ in the integrand, we proceed to the second limit $s_3\to 0$. We find 6 regions, but again only one gives the term $s_3^0$. It corresponds to rescaling 
\begin{equation}
    x_{2,4,5,7,8}\to s_3^{-1} x_{2,4,5,7,8}\,.
\end{equation}
The three remaining limits are done similarly, for each asymptotics we have only one region. The resulting rescaling reads
\begin{equation}
    x_{1}\to \frac{s_4}{s_2} x_1\, \quad
    x_{2,4,5}\to \frac{s_4s_5}{s_2} x_{2,4,5}\, \quad
    x_7\to \frac{s_5}{s_2} x_7\,.
\end{equation}
As a result we obtain a much more accessible expression for this contribution:
\begin{multline}
    PB_{43}^{\text{dci}}=\Gamma (3-\epsilon )
    \intop_{\mathbb{R}_+^8} dx_1\ldots dx_8 x_{24} x_{78}  x_1^{\a_1} x_2^{\a_2} x_3^{-\a_{12\bar 3}} x_5^{\a_4} x_6^{\a_5} x_7^{\a_6} x_8^{-\a_{\bar 3456}} x_4^{-\a_3-2\e }
    \\
    \times\frac{\Big(x_8 \left(x_4x_{125}+\hat{x}_{25}\right)+x_7 \left(x_4x_{136}+x_6x_{24}\right)+\left(x_{36}+1\right) \left(\hat{x}_{24}+\hat{x}_{25}+\hat{x}_{45}\right)\Big)^{-3+\epsilon}}{\Gamma \left(1+\a_1,1+\a_2,1-\a_{12\bar 3},1+\a_4,1+\a_5,1-\a_{\bar 3456},1+\a_6,1+\a_{\bar 3}-2\e,\a_{\bar 3}-1-\epsilon\right)}
\end{multline}
After performing subsequent rescalings
\begin{equation}
    x_{4,5,7,8}\to x_2 x_{4,5,7,8}, \quad
    x_7\to \frac{x_4+x_5+x_4x_5}{1+x_4} x_7
\end{equation}
and taking the integrals in the order $x_1, x_3, x_6, x_2, x_8, x_5, x_4, x_7$, we obtain the contribution of the hard region:
\begin{equation}
    PB_{43}^{\text{dci}}=
    \frac{\Gamma \left(1-\alpha _{12\bar34},1+\alpha _{2\bar5},-\alpha _{2\bar346},\alpha _{12\bar346},-\alpha _{12}-\epsilon ,\alpha _{2\bar3}-\epsilon ,\alpha _{\bar345}-\epsilon ,1+\epsilon +\alpha _3\right)}{\Gamma \left(\alpha _1,1+\alpha _2,1-\alpha _{12\bar3},1-\alpha _{\bar3456},1+\alpha _6,1-2 \epsilon -\alpha _3,1-\epsilon +\alpha _{2\bar34},1-\epsilon -\alpha _{12\bar5}\right)}
\end{equation}
We present the contributions of all regions in terms of $\Gamma$-functions for generic $\e, \a_1,\ldots,\a_6$ in the ancillary files, see Section \ref{sec:ancillary}. Summing up all contributions and expanding in $\a_i$ and $\e$, we obtain our final result for the slightly off-shell pentabox integral:
\begin{multline}\label{eq:pentabox_l}
    PB=\tfrac{1}{2} (l_3+l_4-l_1) (l_2^3+2 l_1 l_2^2+l_3 l_2^2-l_4 l_2^2-l_5 l_2^2-l_3^2 l_2-l_5^2 l_2+4 l_1 l_3 l_2\\
    -2 l_3 l_4 l_2-4 l_1 l_5 l_2+2 l_3 l_5 l_2+2 l_4 l_5 l_2+l_5^3-l_1 l_3^2-l_1 l_4^2+l_3 l_4^2+2 l_1 l_5^2\\
    -l_3 l_5^2+l_4 l_5^2+2 l_1^2 l_3+2 l_1^2 l_4+l_3^2 l_4-4 l_1 l_3 l_4-l_4^2 l_5+4 l_1 l_4 l_5-2 l_3 l_4 l_5)\\
    +\tfrac{1}{2} (3 l_3^2-4 l_1 l_3-2 l_2 l_3+8 l_4 l_3+3 l_4^2+2 l_1 l_2-4 l_1 l_4+2 l_1 l_5+4 l_2 l_5-2 l_4 l_5)\zeta _2 \\
    +(4 l_1+l_2-3 l_3-3 l_4+l_5)\zeta _3 +5 \zeta _4 +O(u_i),
\end{multline}
where $l_i=\log u_i$. This formula can be further simplified if we introduce slightly different conformal variables $U_i$ and the corresponding logarithms $L_i=\log U_i$. These variables are related to $u_i$ via
\begin{equation}
    U_i=\frac{u_{i-2}u_{i+2}}{u_i},\qquad
    u_i=\frac{U_{i-1}U_{i+1}}{U_i}.
\end{equation}
Then we have
\begin{multline}\label{eq:pentabox}
    PB=\tfrac{1}{2} L_1 \left(L_3 L_2^2+2 L_3 L_4 L_2+L_4 L_5^2+2 L_3 L_4 L_5\right)
    +\tfrac{1}{2} \big(4 L_1 L_2+4 L_1 L_5+4 L_3 L_4+2 L_1 L_3
    \\
    +2 L_1 L_4-2 L_4 L_2-2 L_3 L_5-L_2^2-L_5^2\big)\zeta _2
    +\left(L_3+L_4-2 L_1\right) \zeta_3+5 \zeta _4 +O(U_i)
\end{multline}
For the single-scale limit, $s_i=Q^2$, we find, in agreement with Ref. \cite{Bork:2022vat}, 
\begin{equation}\label{PBintSymPoint}
PB\Big{|}_{s_i=Q^2}=3L^4+5\zeta_2L^2+5\zeta_4+O (u),
\end{equation}
with $L=\log u=\log U$ and $u=U=m^2/Q^2$.

Our result \eqref{eq:pentabox} has a remarkably simple form as compared with that of recent paper \cite{Belitsky:2025sin}. However, we have successfully checked the numerical agreement of those two results.

\section{More examples}

\begin{figure}
	\centering
	\includegraphics[width=0.7\linewidth]{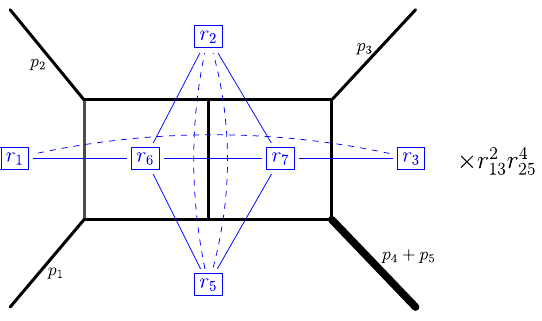}\\
	\includegraphics[width=0.7\linewidth]{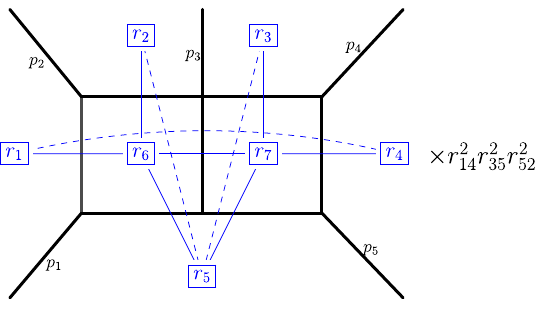}
	\caption{DCI off-shell double box $DB_{\text{off}}$ (top) and double box with five legs $DB_5$ (bottom).}
	\label{fig:more}
\end{figure}

In addition to the pentabox integral, we have also considered the integrals $DB_{\text{off}}$ and $DB_{5}$ depicted in Fig. \ref{fig:more}. The most generic DCI regularization of the first integral is given by
\begin{multline}
	DB_{\text{off}}^{\text{dci}}=DB_{\text{off}}^{\text{dci}}(\e,\a_1,\ldots, \a_6)\\
	=\int \frac{d^dr_6}{\pi^{d/2}}\frac{d^dr_7}{\pi^{d/2}}
	\frac{r_{14}^{2\left(\a_6-\a_2\right)}r_{24}^{2\left(\a_2-\a_6\right)}r_{35}^{2\left(\a_3-\a_5\right)}r_{25}^{2\left(\a_{56}+2\right)}r_{13}^{2\left(\a_{12}-\a_6+1\right)}}{r_{61}^{2\left(1+\a_1\right)}r_{62}^{2\left(1+\a_2\right)}r_{65}^{2\left(1+\a_3\right)}r_{67}^{2\left(1-\a_{123}-2\e\right)}r_{72}^{2\left(1+\a_5\right)}r_{75}^{2\left(1+\a_6\right)}r_{73}^{2\left(1+\a_{123}-\a_{56}\right)}}
\end{multline}
and of the second integral by
\begin{multline}
	DB_{5}^{\text{dci}}=DB_{5}^{\text{dci}}(\e,\a_1,\ldots, \a_5)\\
	=\int \frac{d^dr_6}{\pi^{d/2}}\frac{d^dr_7}{\pi^{d/2}}
	\frac{r_{35}^{2\left(1+\a_3\right)}r_{25}^{2\left(1+\a_4\right)}r_{24}^{2\left(\a_2-\a_4\right)}r_{13}^{2\left(\a_5-\a_3\right)}r_{14}^{2\left(\a_{13}-\a_5+1\right)}}{r_{61}^{2\left(1+\a_1\right)}r_{62}^{2\left(1+\a_2\right)}r_{65}^{2\left(1+\a_3\right)}r_{67}^{2\left(1-\a_{123}-2\e\right)}r_{75}^{2\left(1+\a_4\right)}r_{73}^{2\left(1+\a_5\right)}r_{74}^{2\left(1+\a_{123}-\a_{45}\right)}}
\end{multline}
We find 18 and 29 different regions for the first and second integral, respectively. 

Applying the same technique as in the previous section, we obtain the following results
\begin{equation}\label{eq:doublebox_off}
DB_{\text{off}} = \tfrac{1}{4} L_4^2 \left(L_3+L_5\right)^2+\tfrac{1}{2}  \left(L_3^2+4 L_4 L_3+2 L_5 L_3+L_4^2+L_5^2+4 L_4 L_5\right)\zeta _2+\tfrac{21 }{2}\zeta _4+O(U_i)
\end{equation}
and
\begin{multline}\label{eq:doublebox_5}
	DB_{5} = \tfrac{1}{4} L_2 L_3 \left(2 L_1+L_3\right) \left(2 L_4+L_2\right) +\left(2 L_1 L_2+2 L_3 L_4+2 L_2 L_3+L_1 L_4\right)\zeta _2 \\
    - \left(L_1+L_4\right)\zeta _3+\tfrac{31 }{4}\zeta _4
	+O(U_i),
\end{multline}

We have checked that Eq. \eqref{eq:doublebox_off} is in agreement with the direct expansion of Davydychev \& Usyukina result \cite{Usyukina:1993ch}. The symmetric limit of  \eqref{eq:doublebox_5} agrees with Ref. \cite{Bork:2022vat}

\section{Ancillary files}\label{sec:ancillary}

We attach the following files to this paper
\begin{itemize}
    \item \texttt{doublebox1of/region*.m}, \texttt{doublebox5/region*.m}, \texttt{pentabox/region*.m} --- contribution of each region exact in all regularization parameters for $DB_{\text{off}}^{\text{dci}}$, $DB_{5}^{\text{dci}}$, and $PB^{\text{dci}}$, respectively.
    \item \texttt{doublebox1of/doublebox1of}, \texttt{doublebox5/doublebox5}, \texttt{pentabox/pentabox} --- final results \eqref{eq:doublebox_off}, \eqref{eq:doublebox_5}, \eqref{eq:pentabox} for $DB_{\text{off}}$, $DB_{5}$, and $PB$, respectively.
    \item \texttt{5points\_2loops.nb} --- demonstration notebook.
\end{itemize}

\section{Conclusion and Outlook}

In the present paper we have introduced a new approach to the calculation of slightly off-shell dual conformal integrals. The approach is based on the method of regions and the regularization which allows one to separate the contributions of different regions in a way compatible with dual conformal symmetry. This regularization is a constrained combination of dimensional and analytic regularizations which preserves the dual conformal invariance of the original integral. We observe that this approach drastically simplifies the calculation of each contribution as compared to conventional approach based on the dimensional regularization alone. We support this observation by the calculation of the two-loop dual conformal pentabox integral, Eq. \eqref{eq:pentabox}, and two simpler integrals, Eqs. \eqref{eq:doublebox_off}  and \eqref{eq:doublebox_5}, in slightly off-shell kinematic. The results for these examples are expressed in terms of logarithms. We have performed a number of cross checks of the obtained results, including the numerical comparison of the pentabox integral with the recent result of Ref. \cite{Belitsky:2025sin} and found perfect agreement.

Let us speculate about possible applications and development of the introduced approach. First, a natural application of the same approach would be the calculation of five-point three loop dual conformal integrals. As the approach is based on explicit direct integration rather than IBP reduction, we do not expect an explosive growth of complexity when going to 3 loops. Another interesting direction of further development would be the calculation of six-point two loop slightly off-shell dual conformal integrals. Again, one could use DCI regularization to separate the contribution of different regions, but now each contribution non-trivially depends on the three cross-ratios made of the invariants of massless amplitude. Therefore, one would benefit from deriving the differential equations for the contributions as functions of those three cross ratios. However, the conventional approach based on IBP reduction in the context of DCI integrals has the same drawback as the dimensional regularization: it breaks the symmetry. One could think of the approach analogous to the conventional IBP reduction, but with the replacement of Lorentz invariance with the conformal invariance. Within this hypothetic approach, one would determine a family of dual conformal integrals and a sufficiently large set of first-order differential operators with respect to integration variables, which generates a number of DCI IBP relations sufficient for the reduction. We are currently actively investigating such perspectives.

\acknowledgments
We are grateful to A. Belitsky and V. Smirnov for stimulating discussions and interest to this work.
\bibliographystyle{JHEP}
\bibliography{DCIreg}
\end{document}